\ifx\empty\selectedoptions
  \def\selectedoptions{final}
\fi

\documentclass[final]{aipproc}

\SetInternalRegister\hbadness{8000} 

\newcommand\doingARLO[2][]{%
  \ifx\mmref\undefined #1\else #2\fi
}
  
\layoutstyle{6x9}

\begin{document}

\title{Fundamental Symmetries of the Early Universe and the Precision Frontier}


\author{Michael J. Ramsey-Musolf}{
  address={Department of Physics, University of Wisconsin-Madison, Madison, WI 53706 USA\\ and \\ 
  Kellogg Radiation Laboratory, California Institute of Technology, Pasadena, CA 91125 USA},
  email={mjrm@physics.wisc.edu},
}

\copyrightyear  {2001}

\begin{abstract}
The search for the next Standard Model of fundamental interactions is being carried out at two frontiers: the high energy frontier involving the Tevatron and Large Hadron Collider, and the high precision frontier where the focus is largely on low energy experiments. I discuss the unique and powerful window on new physics provided by the precision frontier and its complementarity to the information we hope to gain from present and future colliders.
\end{abstract}

\date{\today}

\maketitle

\section{Introduction}
During the coming decade, we hope to discover the larger theory in which the highly successful Standard Model (SM) is embedded. There exist strong reasons to believe that one consequence of this theory -- what I will call the \lq\lq new Standard Model\rq\rq\ (NSM) -- will involve new TeV scale particles that can be discovered at the Large Hadron Collider (LHC). An equally important way to search for the NSM is to carry out ultra-precise measurements at relatively low-energies -- measurements that are either sensitive to tiny deviations from SM predictions or to phenomena that are highly-suppressed or forbidden by SM symmetries. In this talk, I will try to illustrate how studies at the precision frontier complement collider searches and how they can, in some cases, probe scales far beyond those accessible at the Tevatron or LHC.

Howie Baer has already given a nice overview of the motivation for thinking beyond the SM. Let me simply highlight a few of the unsolved puzzles for which the precision frontier might hold important clues:  What is the origin of matter (both visible and dark) ?  What is the dark energy and what is the nature of the dark sector ?
What is the origin of the dimensionful parameters of the SM ($m_{q,\nu}$,  $G_F$, $\Lambda_\mathrm{QCD}$,...) and why are they stable against quantum corrections ?
 What are the discrete symmetries of the early universe (P, CP, T, B, L,...) ? When and how were they broken ?

In the remainder of the talk, I will discuss how precision frontier studies may shed new light on these questions. First, however, I will make a disclaimer: due to time and space limitations, I can only treat a few illustrative examples rather than give a comprehensive survey. I will leave out many important experimental efforts and theoretical developments, and will concentrate on three areas: searches for the permanent electric dipole moments (EDMs) of the electron, neutron, neutral atoms, and nuclei; searches for the neutrinoless double beta decay ($0\nu\beta\beta$) of heavy nuclei; and precision tests of parity-violating electron scattering and of Einstein's weak equivalence principle. For a discussion of other topics, please refer to my recent reviews\cite{Erler:2004cx,RamseyMusolf:2006vr}.

\section{Electric Dipole Moment Searches}
As Norval Fortson has already discussed, the most sensitive test of CP symmetry in flavor conserving interactions is the search for a static EDM. In the SM, one expects a non-vanishing EDM due to the CP-violation (CPV) in the CKM matrix, but the expected magnitude lies well below the current and prospective sensitivity of EDM searches, as illustrated in Table \ref{tab:edm}.

The suppression of the SM CKM-based expectations occurs because this origin of EDMs requires two flavor changing vertices and is known to start off, for an elementary fermion, at three-loop order. On the other hand, the possible size of an EDM arising from another source could be as large as the present limits. In the case of CPV associated with the QCD $\theta$-term, the naive theoretical prejudice is that it should lead to an EDM of the neutron that is roughly $10^{10}$ times larger than the present limit. The stringent $d_n$ limit presents a puzzle as to why the corresponding parameter ${\bar\theta}$ is so tiny. One solution -- spontaneously broken Peccei-Quinn symmetry -- may help explain the dark matter mystery in the guise of the associated axion. If so, it may be that the observation of a neutron or mercury EDM associated with ${\bar\theta}$ is  just around the corner.

On the other hand, if an observable EDM is associated with new physics rather than the massless gluons of QCD, then the effects of the new CPV will scale as $1/M^2 \times \sin\phi_\mathrm{CPV}$, where $M$ is the mass scale associated with the new particles and $\phi_\mathrm{CPV}$ is a CPV phase. Assuming that the corresponding EDM arises at one-loop order, then using naive dimensional analysis we see that the present limits imply that these experiments are probing either very high mass scales or very small phases:
\begin{equation}
\sin\phi_\mathrm{CPV} \sim 1 \rightarrow M > 5000\ \mathrm{GeV} \quad \mathrm {or}\quad M < 500 \ \mathrm{GeV}  \rightarrow \sin\phi_\mathrm{CPV}  < 10^{-2}\ \ \ .
\end{equation}
Future experiments hope to improve the sensitivities by factors of as much as 100 over the next several years, implying probes of physics at the 50 TeV scale or phases at the $10^{-4}$ level. Either way, these experiments will have a reach extending well beyond that of the LHC and will search for CPV effects that are too small to be pulled out of the hadronic collider environment. 

\begin{table}
\begin{tabular}{lllll}
\hline
	\tablehead{1}{l}{b}{Particle}
	& \tablehead{1}{l}{b}{System}
  & \tablehead{1}{l}{b}{EDM Limit\\ in $e$-cm}
  & \tablehead{1}{l}{b}{SM CKM \\ Prediction}
  & \tablehead{1}{l}{b}{New \\ CPV ?}   \\
\hline
e & Tl atom & $1.9\times 10^{-27}$ &  $10^{-38}$   & $10^{-27}$ \\
n & UCN & $2.9\times 10^{-26}$ &  $10^{-31}$   & $10^{-26}$ \\
$^{199}$Hg &atom cell  & $3.1\times 10^{-29}$ &  $10^{-33}$   & $10^{-29}$ \\
\hline
\end{tabular}
\caption{Current limits on the EDM of the electron, neutron, and $^{199}$Hg atom, along with expectations based on the known CPV associated with the CKM matrix and of possible new sources of CPV. \lq\lq UCN\rq\rq\ denotes ultracold neutrons. Courtesy of C.-P. Liu.}
\label{tab:edm}
\end{table}

EDM searches are also important for addressing one of the outstanding problems in cosmology: the origin of the visible matter of the universe. The inflationary paradigm suggests that the universe was likely to be matter-antimatter symmetric at the end of inflation, in which case, the particle physics of the subsequently evolving cosmos would have to be responsible for generating the asymmetry. Four decades ago, Sakharov identified the essential ingredients needed for this to occur\cite{Sakharov:1967dj}: violation of baryon number conservation; violation of both C and CP conservation; and a departure from equilibrium dynamics, assuming that CPT is conserved. These criteria, as applied to the possibility of electroweak baryogenesis (EWB), are listed in Table~ \ref{tab:baryon}. In EWB, a first order phase transition occurs during which bubbles of broken electroweak symmetry nucleate in a symmetric background. CPV interactions at the bubble wall lead to a net density of left-handed fermions that diffuses ahead of the wall, where electroweak sphalerons convert it into baryon number. The expanding bubbles capture this baryon number by quenching the sphalerons. This process will generate the observed baryon-to-photon ratio provided that (1) sufficiently large CPV asymmetries are generated during the transition and (2) the transition is strongly first order in order to quench the sphalerons after baryon number is created. 

As indicated in Table~\ref{tab:baryon},  EWB is not viable in the SM even though it contains, in principle, all the necessary ingredients. The effects of CKM CPV are too weak to generate the large left-handed particle density needed to drive the sphalerons. Moreover, it is now known that electroweak symmetry breaking in the SM proceeds via a cross-over transition rather than a first order or  even second order transition because the LEP lower bound on the SM Higgs mass is too high to allow a phase transition to occur. Consequently, if EWB is responsible for the observed baryon asymmetry, new electroweak physics is needed.

One of the most attractive possibilities is supersymmetry. In the Minimal Supersymmetric Standard Model (MSSM), loop effects associated with scalar superpartners of the right handed (RH) top quark can lead to a strong first order phase transition compatible with the LEP bounds if the RH stop mass is less than about 125 GeV \cite{Carena:2008vj}. In addition, the MSSM contains a plethora of new CPV phases whose effects are not suppressed by light quark Yukawa couplings as in the SM. Nevertheless, EDM limits place severe constraints on the size of these phases and, thus, on the viability of supersymmetric EWB. In a scenario where the one-loop contributions dominate, EWB in the MSSM is marginally viable, if at all. For superpartners with masses below a TeV, the generic bounds on the phases of $\sim 10^{-2}$ preclude the production of large CPV asymmetries during an electroweak phase transition. On the other hand, taking the superpartner masses to be several TeV in order to allow for larger CPV phases implies that they would decouple from the plasma during the electroweak phase transition, thereby having no impact on the production of left-handed fermion number density.

\begin{table}
\begin{tabular}{lllll}
\hline
	\tablehead{1}{l}{b}{Criterion}
	& \tablehead{1}{l}{b}{Standard\\ Model?}
	& \tablehead{1}{l}{b}{Origin}
	& \tablehead{1}{l}{b}{New Physics\\ Needed ?}
  & \tablehead{1}{l}{b}{Experimental \\ Probe}   \\
\hline
B Violation & Yes & Weak Sphalerons &  No & --   \\
C  and CP Violation &  Yes & CKM Matrix &  Yes   & EDM \\
Noneq dynamics  & No &  Scalar Potential   & Yes & LHC  \\
\hline
\end{tabular}
\caption{Sakharov criteria as applied to electroweak baryogenesis.}
\label{tab:baryon}
\end{table}

The way around this seeming conundrum is to consider a scenario in which the scalar superpartners of the first and second generation fermions are heavy while the gauginos and higgsinos remain light. In this case, one can suppress the one-loop EDMs and in principle allow for larger CPV phases. Moreover, the light gauginos and higgsinos remain active during the phase transition, and it is their CPV interactions that could ultimately drive the MSSM EWB. One still has to content with EDM limits, however, because the light gauginos and higgsinos lead to important two-loop EDM contributions through the well-known \lq\lq Barr-Zee\rq\rq\ diagrams. 

Recently, we completed a computation of these two-loop graphs\cite{Li:2008kz,Li:2008ez} and found that the resulting EDM limits on the phases relevant to EWB can still be relatively severe. In particular, electron and neutron EDM limits imply that the phase $\phi_2\equiv\mathrm{Arg}(\mu M_2 b^\ast)$ associated with the supersymmetric $\mu$ parameter, the soft wino mass parameter $M_2$, and the soft Higgs mass parameter $b$ is too small to lead to successful EWB. On the other hand, two-loop EDMs depend much less strongly (by about a factor of fifty) on the phase $\phi_1\equiv\mathrm{Arg}(\mu M_1 b^\ast)$ involving the soft bino mass parameter. Consequently, the most viable path to MSSM EWB involves CPV bino-higgsino interactions during the phase transition, while those involving winos are unlikely to play a significant role. Moreover, EDM limits imply that the SUSY-breaking mechanism must be non-universal -- in contrast to, say, mSUGRA -- so that $\phi_1$ and $\phi_2$ are independent parameters. 

It is interesting to speculate that the up-coming EDM searches will discover a non-zero result that -- in the context of the MSSM -- is consistent with a large $\phi_1$ phase needed for EWB. Would we then conclude that EWB is responsible for the observed baryon asymmetry? Not necessarily. As emphasized by the work of Carena {\em et al}\cite{Carena:2008vj}, one would also need to discover a light RH stop at the Tevatron or LHC to conclude that there was a strong first order phase transition. Even then, however, we will require additional information. In particular, the results from the muon $g-2$ experiment point to a fairly large value for $\tan\beta$, implying that the dynamics of the superpartners of the bottom quarks,  tau leptons, and their superpartners may not be negligible during the phase transition as previously assumed. Recently, we showed that the Yukawa interactions involving these particles can lead to a quenching of the baryon asymmetry for moderate to large $\tan\beta$\cite{Chung:2008aya}. This effect can be particularly pronounced if the RH top and bottom squark masses are not too different. Thus, we will not only need to learn from the colliders that the RH stop is light but that the RH sbottom and stau are not so light. In short, it will ultimately take information from both the EDMs and the colliders to determine the viability of supersymmetric EWB.

\section{Neutrinoless Double Beta Decay}
\label{sec:DBD}

If the combination of EDMs and collider searches ultimately rule out EWB, then the favored baryogenesis scenario will undoubtely be leptogenesis. The standard leptogenesis scenario requires lepton number violation, associated with the existence of a Majorana mass term for the neutrinos. There also exist Dirac leptogenesis scenarios, which do not require lepton number violation. However, the see-saw paradigm for explaining the tiny scale of $m_\nu$ is so theoretically appealing that most people concentrate on the standard leptogenesis picture that implies a Majorana mass. 

While it is possible that LHC experiments may discover lepton number violation through the observation of same-sign diplepton pairs, the discovery potential is model-dependent. The more generic way to look for the violation of this symmetry is to search for $0\nu\beta\beta$. Assuming that the decay proceeds by the exchange of a virtual light Majorana neutrino, then the rate is proportional to the square of the {\em effective mass} $m_{\beta\beta}$ that is not in general identical with the mass of the lightest neutrino, $m_1$. The relationship between the two involves the two possible Majorana phases and depends on the neutrino mass hierarchy. For the quasi-degenerate or normal hierarchies, $m_{\beta\beta}\sim m_1$, while for the inverted hierarchy, $m_{\beta\beta}$ can be about an order of magnitude larger than $m_1$. The present generation of $0\nu\beta\beta$ experiments will be able to probe for a signal associated with $m_{\beta\beta}$ of tens of meV, corresponding to the inverted hierarchy range. The normal hierarchy implies a much smaller effective mass while in the quasi-degenerate hierarchy it is much larger. If nature is kind to the experimentalists working at this portion of the precision frontier, the the hierarchy will be either quasi-degenerate or inverted, a convincing signal will be observed, and we will know that neutrinos are Majorana fermions. 

It is also interesting to ask what the absence of an observation might mean. Could it tell us that neutrinos are Dirac particles? Possibly so, but this scenario requires input from other neutrino experiments. In particular, the KATRIN experiment is measuring the $\beta$ spectrum of tritium $\beta$-decay, looking for behavior at the endpoint that might indicate a non-zero value of $m_1$. Given the sensitivity of this experiment, if an endpoint deviation is observed, it would tell us the absolute scale of neutrino mass (something oscillation experiments alone cannot do) and imply that the spectrum is quasi-degenerate. If neutrinos are Majorana particles, one would then expect a non-zero result in the present $0\nu\beta\beta$ experiments. The absence of an observation would presumably imply that neutrinos are Dirac. If KATRIN obtains a null result, future long baseline experiments may indicate -- if we are fortuante -- an inverted hierarchy. If so, one would again expect a signal in the $0\nu\beta\beta$ searches and conclude from the absence of observation that we have light Dirac neutrinos.

As intriguing a possibility as this may be, there exists an important loophole. It is possible that the mechanism responsible for $0\nu\beta\beta$ is not the exchange of a light Majorana neutrino but rather the exchange of one or more heavy particles that entail lepton number violating (LNV) interactions (see, {\em e.g.}, Ref.~\cite{Cirigliano:2004tc} and references therein) . In R parity-violating (RPV) SUSY, for example, the decay can proceed via the LNV $LQ{\bar D}$ operators and the exchange of the Majorana gluino. Similarly, left-right symmetric models with heavy right-handed Majorana neutrinos can lead to the same situation. If the mass scale of the heavy particles involved in the exchange is at the TeV scale, then the corresponding $0\nu\beta\beta$ amplitude can be comparable to the expected amplitude associated with light Majorana neutrino exchange. Until we are certain as to the dominant mechanism, we cannot conclusively interpret either a signal or limit without making an additional assumption as to the mechanism. 

To illustrate, suppose that a $0\nu\beta\beta$ experiment observes a signal consistent with the inverted hierarchy  but that the actual mechanism for the decay is RPV SUSY with gluino exchange. One would be tempted to conclude that $m_{\beta\beta}$ is of order a few tens of meV. However, the same RPV interactions that enter the $ue^+ {\tilde d}$ vertices in the decay will also generate a one-loop Majorana neutrino mass that is roughly an order of magnitude smaller, corresponding to the normal hierarchy. In this instance, lack of knowledge of the mechanism could lead to a the wrong conclusion about the absolute scale of neutrino mass. On the bright side, the observation $0\nu\beta\beta$ implies without question that neutrinos are Majorana particles, as shown by Schecter and Valle\cite{Schechter:1981bd}. We theorists simply would not be able to say much more without additional information.

This \lq\lq mechanism problem\rq\rq\ is discussed extensively by Petr Vogel in his contribution to these proceedings, and I refer the reader to that discussion for additional details. As he points out, it may be possible to use information from experimental probes of charged lepton flavor violation (CLFV), since in many scenarios that have low-scale LNV one finds a corresponding expectation of observable CLFV. Nevertheless, determining the decay mechanism remains an important, open theoretical problem -- along side the more familiar problem of computing the nuclear matrix elements -- and one that should be pursued in tandem with the impressive experimental efforts.

\section{Precision Tests}

The searches for EDMs and $0\nu\beta\beta$ exemplify the precision frontier studies involving processes that are highly suppressed or forbidden in the SM. Considerable experimental and theoretical efforts are also being devoted to studies of observables that are not forbidden, such as the muon anomalous magnetic moment. Any convincing deviation from a SM prediction for such an observable could point to new physics. More generally the pattern of deviations, or lack thereof, from a variety of these \lq\lq precision tests\rq\rq\ provide important guidance into the nature of the NSM. Among the most precise tests are those involving studies of weak decays, parity-violating electron scattering, muon properties and interactions, and gravitational interactions. Due to space limitations, I will only comment on two of these: parity-violating electron scattering and tests of Einstein's weak equivalence principle (WEP). I refer the reader to the various contributions treating neutron and nuclear $\beta$-decay, pion leptonic decays, the muon anomalous magnetic moment for details on these important classes of precision tests.

Turning first to parity-violating electron scattering (PVES), the frontier during the next decade involves scattering of longitudinally polarized electrons from fixed targets, such as protons or electrons in hydrogen. One measures the PV asymmetry for scattering involving positive and negative helicity electrons (for reviews, see Refs.~\cite{Musolf:1993tb,Beck:2001dz,Beise:2004py}):
\begin{equation}
A_\mathrm{PV} = \frac{N_+-N_-}{N_++N_-} = \frac{G_F Q^2}{4\sqrt{2}\pi\alpha}\left[Q_W+F(Q^2)\right]\ \ \ ,
\end{equation}
where $N_+$ ($N_-$) are the number of counts for positive (negative) helicity electrons. The quantity of interest for new physics is the weak charge of the target, $Q_W$, while the case of hadronic/nuclear targets, the form factor $F(Q^2)$ has been under intensive scrutiny for two decades as a way of probing the strange quark contributions to the nucleon's electromagnetic structure. The weak charge and $F(Q^2)$ can be experimentally separated by exploiting the $Q^2$-dependence of the latter. 

The weak charges of the proton and electron are particularly interesting as a window on new physics, as both are proportional to $1-4\sin^2\theta_W$ (at tree level). Due to the near cancellation between the two terms in this expression, the SM predictions for $Q_W^{e,p}$ are suppressed, leading to a relatively enhanced sensitivity to new physics. For the same reason, a precise measurement of $Q_W^{e,p}$ can provide a precise determination of the weak mixing angle. This feature motivated the recently completed E158 PV Moller scattering experiment at SLAC and provides part of the rationale for the upcoming Q-Weak experiment and future PV Moller experiment at Jefferson Laboratory. A determination of $\sin^2\theta_W$ in these experiments is interesting, even in light of the per mil accuracy of $\sin^2\theta_W$ determinations at LEP and the SLC, because the SM predicts that the weak mixing angle runs with energy scale. Apart from looking for new physics, the low-energy PVES experiments provide a test of this predicted running, as discussed in detail in Refs.~\cite{Czarnecki:2000ic,Erler:2004in} . 

\begin{figure}
  \resizebox{20pc}{!}{\includegraphics{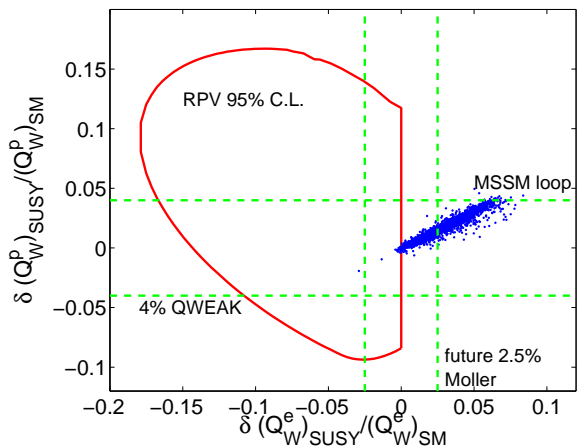}}
\caption{\label{fig:susy} Supersymmetric contributions to the weak charges of the proton (vertical axis) and electron (horizontal axis) relative to SM predictions. Blue dots indicate loop corrections for randomly chosen values of the MSSM parameters. Interior of the red region indicates shifts due to RPV interactions including constraints from a global fit to precision observables. Horizontal and vertical dashed green bands indicate proposed one sigma sensitivities of the Jefferson Lab Q-Weak and PV Moller experiments. Figure courtesy of S. Su. }
\end{figure}

To illustrate the new physics sensitivity of the future PVES experiments, I will return to the MSSM. In this scenario, $Q_W^{e,p}$ can deviate from the SM expectations (taking the Z-pole weak mixing angle as input and using its SM running) due to two effects: loop corrections involving superpartners and tree-level contributions to the PV amplitudes arising from RPV interactions. The relative impact of these two effects on the weak charges is illustrated in Figure \ref{fig:susy}, based on the work of Ref.~\cite{Kurylov:2003zh}.

As one can see, the signatures of MSSM loops and RPV interactions on these two weak charges are rather distinctive. If SUSY is discovered at the LHC, the combination of these PVES measurements may provide an interesting \lq\lq diagnostic tool\rq\rq\  for determining whether or not one has an R parity conserving or violating scenario. As indicated above, the presence of RPV interactions in this context would imply a Majorana mass for the light neutrinos. Moreover, it would preclude a neutralino dark matter candidate in the MSSM. The central value of the E158 results lies in the RPV-favored region, but the error bars are too large to be anything more than suggestive. On the other hand, the muon $g-2$ favors light superpartners with large $\tan\beta$ -- a region of MSSM parameter space that favors the larger MSSM loop corrections to the weak charges. Against this background, it will be interesting to see if the PVES results favor one direction or the other -- or an altogether different new physics scenario. Either way, the information these exquisitely precise measurements can provide will be complementary to what we may learn from the LHC. 

As a second example of precision tests, I will consider something of a more speculative nature involving the dark sector. Given the richness of interactions and particle species that make up the visible sector and the $\sim$five times larger abundance of dark matter compared to visible matter, it is not  unreasonable to imagine that the dark sector consists of more than one particle species and that there exist new interactions that reside primarily in the dark sector. If so, it is interesting to consider how one might know. 

One possibility is that there exists a new, long-range non-gravitational force between dark matter particles mediated by the exchange of an ultralight scalar that would lead to a violation of Einstein's weak equivalence principle (WEP) in the dark sector. As discussed by Bovy and Farrar\cite{Bovy:2008gh}, the existence of such a force could alleviate some tensions in the $\Lambda$CDM fit to astrophysical and cosmological observations. Recently, Kesden and Kamionkowski have derived approximate bounds on the strength of a long range dark force from tidal streams in the Sagitarius galaxy\cite{Kesden:2006vz}, while Bean {\em et al} have arrived at limits analyses of the CMB\cite{Bean:2008ac}. Additional constraints can be obtained from terrestrial experiments if the dark matter is not sterile\cite{Carroll:2008ub,Carroll:2009dw,Bovy:2008gh}. In this case, loop effects involving virtual DM particles that couple to SM fermions through weak interactions or Higgs exchanges would induce a coupling of ordinary matter to the ultralight scalar. The result would be violations of the WEP in matter-dark matter and matter-matter interactions. 

Searches for such WEP violation have been carried out with torsion balance experiments, leading to an upper bound of $(0.3\pm 1.8)\times 10^{-13}$ and $(4\pm 7)\times 10^{-5}$ on the differential acceleration of two different test bodies compared to their common gravitational acceleration toward the center of the earth ($\eta_\mathrm{E}$)  and galactic center  ($\eta_\mathrm{DM}$), respectively\cite{Schlamminger:2007ht}. At present, the astrophysical analyses lead to stronger bounds on typical WIMP model parameter space for this scenario than do the torsion balance experiments, but any improvements in the latter would make the latter a more powerful probe. The space based Microscope experiment being carried out by the European Space Agency aims to improve the sensitivity to $\eta_\mathrm{E}$ by a factor of one hundred over current bounds, while the MiniSTEP experiment being considered by NASA and the ESA might achieve an additional three orders of magnitude sensitivity. It will be difficult, if not impossible, to look for consequences of this long range dark force at the LHC. 

\section{Conclusions}

In this talk, I hope to have illustrated how studies at the precision frontier are an essential complement to the LHC in searching for the symmetries of the NSM. Searches for SM-forbidden or suppressed phenomena, such as the EDM or $0\nu\beta\beta$, could reveal violations of fundamental symmetries that are needed to explain the origin of matter, while precision tests such as PVES, weak decays, studies of muon properties, and torsion balance experiments can probe detailed aspects of potential new forces. Overall, there is a rich potential for both discovery and insight at the precision frontier, making for an exciting era at the intersection of particle and nuclear physics with astrophysics, cosmology, and atomic physics. 

\begin{theacknowledgments}
I wish to thank the CIPANP 2009 organizers for their hospitality and the Aspen Center for Physics where part of this manuscript was completed. I also thank my many collaborators in this field, C.-P. Liu for providing input for Table I, and Shufang Su for providing the figure. This work was supported in part by the U.S. Department of Energy contract DE-FG02-08ER41531 and the Wisconsin Alumni Research Foundation. 
\end{theacknowledgments}


\doingARLO[\bibliographystyle{aipproc}]
          {\ifthenelse{\equal{\AIPcitestyleselect}{num}}
             {\bibliographystyle{arlonum}}
             {\bibliographystyle{arlobib}}
          }

\end{document}